\documentclass[cits]{PoS}
\usepackage{amssymb}
\usepackage{wasysym}
\usepackage{graphicx}
\usepackage{fontenc}
\usepackage{times}
\usepackage{mathptmx}
\title{Protosolar Irradiation in the Early Solar System: Clues from Lithium and Boron Isotopes}

\ShortTitle{Early Solar System Irradiation }

\author{\speaker{Ming-Chang Liu}\\
        Department of Terrestrial Magnetism, Carnegie Institution of Washington\\
        Institute of Astronomy and Astrophysics, Academia Sinica\\
        E-mail: \email{mliu@dtm.ciw.edu}}

\author{Larry R. Nittler\\
        Department of Terrestrial Magnetism, Carnegie Institution of Washington\\
        E-mail: \email{lrn@dtm.ciw.edu}}

\author{Conel M. O'D. Alexander\\
        Department of Terrestrial Magnetism, Carnegie Institution of Washington\\
        E-mail: \email{alexande@dtm.ciw.edu}}

\author{Typhoon Lee\\
        Institute of Astronomy and Astrophysics, Academia Sinica\\
        E-mail: \email{typhoon@asiaa.sinica.edu.tw}}

\abstract{We report Li and B isotopic compositions of 10 Spinel-HIBonite spherules (SHIBs) separated from the Murchison meteorite, in order to understand their irradiation history in the early Solar System. The extremely low Be concentrations in SHIBs preclude detection of extinct $^{10}$Be, but instead allow for a search of the original Li and B isotopic ratios of the grains, as these isotopes are sensitive indicators for irradiation. We found that some of the SHIBs carried sub-chondritic $^{7}$Li/$^{6}$Li and supra-chondritic $^{10}$B/$^{11}$B ratios. Considering two possible irradiation scenarios that could have occurred in the early Solar System, irradiation of hibonite solids followed by addition of isotopically normal Li and B seems to be the most plausible explanation for the observed Li and B isotope ratios.}

\FullConference{11th Symposium on Nuclei in the Cosmos, NIC XI\\
		July 19-23, 2010\\
		Heidelberg, Germany}

\begin{document}

\section{Introduction}
Intense irradiation from the proto-Sun in the early Solar System is believed to have occurred from both cosmochemical and astronomical viewpoints. The former involves the prior existence of $^{10}$Be (t$_{1/2}$ = 1.387 My, \cite{MCLIU:Kor2010}) in the solar nebula, which has been successfully inferred in many meteoritic Ca-Al-rich Inclusions (CAIs) through the radiogenic excesses of $^{10}$B \cite{MCLIU:McK2000a, MCLIU:Sug2001a, MCLIU:Mac2003, MCLIU:Cha2006a}. Astronomical evidence comes from intense X-ray emissions in young stellar objects (YSOs), which is a manifestation of strong flare activities, observed by the {\it{Chandra}} space telescope \cite{MCLIU:Fei2002}. The stable isotopes of Li ($^{7}$Li can also be produced by both Big Bang and stellar nucleosynthesis \cite{MCLIU:Pra2007}) and B are primarily produced by irradiation, and the isotopic compositions of the two elements are very sensitive to the spallation energy \cite{MCLIU:Ram96}. The $^{7}$Li/$^{6}$Li production ratio during irradiation by Galactic cosmic rays (GCRs) is $\sim 1-2$, whereas this ratio in the Solar System is around 12 \cite{MCLIU:Ram96, MCLIU:Sei2007}. Therefore, a detailed understanding of the stable Li and B isotopic compositions of the oldest refractory inclusions in the Solar System could help constrain the irradiation processes in the solar nebula.

Hibonite (CaAl$_{12}$O$_{19}$) is considered to be one of the earliest solids in the Solar System from a thermodynamics perspective \cite{MCLIU:Gro72}, and hibonites found in primitive meteorites can be classified into two groups based on the inferred abundances of the short-lived radionuclide $^{26}$Al \cite{MCLIU:Ire88}. The fossil record of $^{10}$Be in CM-chondrite $^{26}$Al-free PLAty-hibonite Crystals (PLACs) has been well established, indicating that PLACs experienced intense protosolar irradiation \cite{MCLIU:Mar2002, MCLIU:Liu2009a, MCLIU:Liu2010}. In contrast, the $^{26}$Al-bearing Spinel-HIBonite spherules (SHIBs) contain too little Be and too much B, so that $^{10}$B excesses are unresolvable by current techniques even if the SHIBs were also irradiated \cite{MCLIU:Liu2009a, MCLIU:Liu2010}. Interestingly, a couple of SHIB samples that carried ``normal'' $^{10}$B/$^{11}$B show sub-chondritic $^{7}$Li/$^{6}$Li ratios, implying that spallogenic Li might have been mixed into the grains \cite{MCLIU:Liu2009a}. Since the Li and B isotopic ratios in low [Be] SHIBs were not affected by the decay of $^{7}$Be and $^{10}$Be, the grains can provide information about the sources of the Li and B, including spallation. In this paper, we report and discuss new data on the stable Li and B isotopic systematics in individual SHIBs, with a goal of better constraining the irradiation history of these refractory solids. 

\section{Mass Spectrometry}
Ten spinel-hibonite spherules, whose size ranged from 20 to 40 $\mu$m, were hand-picked from a Murchison acid residue (Courtesy of A. Davis, Univ. of Chicago), and mounted in epoxy for isotope measurements. The Li and B mass spectrometry was performed on the CAMECA ims-6f ion microprobe at the Carnegie Institution. The polished samples were sputtered with a 22.5 KeV, 18 nA $^{16}$O$^{-}$ primary beam ($\phi \sim 35 \mu$m) to obtain sufficient signals for each of the Li and B isotopes. Mass resolution (M/$\Delta$M) was set at 1500 to separate the peaks of interest from isobaric interferences. To minimize the contribution of surface B contamination, presputtering was applied to the samples until the B signals became steady. Secondary ions were collected with the axial electron multiplier (EM) by cycling the magnetic field through the mass sequence 5.9, $^{6}$Li, $^{7}$Li, $^{9}$Be, $^{10}$B, $^{11}$B and $^{27}$Al$^{2+}$. The deadtime effect of the EM was negligible because secondary ion signals (except $^{27}$Al$^{2+}$, which is only used for estimating Li, Be and B concentrations) were less than a few thousand counts per second. The instrumental mass fractionation and relative sensitivity factors were determined and corrected for by measuring an NBS 612 glass ($^{10}$B/$^{11}$B = 0.2469 \cite{MCLIU:Kas2001}; $^{7}$Li/$^{6}$Li = 12.3918 [Dalpe et al. private communication]). GCR-produced Li and B isotopes in the hibonite grains produced while the host meteorite was a small object in interplanetary space were estimated by adopting an exposure age of 1.8 Myr for the Murchison meteorite \cite{MCLIU:Her97}, but were found to be negligible due to high concentrations of Li and B in the samples.

\section{Lithium and Boron Isotopic Results}
The Li and B isotopic compositions and concentrations of the 10 analyzed hibonite grains are summarized in Table 1. The $^{7}$Li/$^{6}$Li ratios of the samples range from 11.57 to 12.36, and the total range ($\sim 6.5\%$) is larger than in any other meteoritic material. In contrast, the boron isotopic compositions in the grains are chondritic ($^{10}$B/$^{11}$B = 0.2481 \cite{MCLIU:Zha96}) within errors, except for two grains (mt3-S11 and mt3-S15) that deviate from chondritic at a 3$\sigma$ level. 

\begin{table}[h]
\begin{center}
{\small
\begin{tabular}{cccccccc}
\hline
{Sample} & {$^{9}$Be/$^{6}$Li ($\pm$2$\sigma$)} & {$^{7}$Li/$^{6}$Li ($\pm$2$\sigma$)} & {$^{9}$Be/$^{11}$B ($\pm$2$\sigma$)} & {$^{10}$B/$^{11}$B ($\pm$2$\sigma$)} & [Li] & [Be] & [B]\\
\hline
mt3-S5 & 0.16$\pm$0.01 & 11.57$\pm$0.08 & 0.003$\pm$0.001 & 0.2503$\pm$0.0018 & 1448 & 24 & 10686\\
mt3-S9 & 0.09$\pm$0.03 & 11.69$\pm$0.05 & 0.013$\pm$0.001 & 0.2494$\pm$0.0027 & 5601 & 51 & 5905\\
mt3-S10 & 0.05$\pm$0.01 & 11.95$\pm$0.03 & 0.005$\pm$0.001 & 0.2496$\pm$0.0015 & 9777 & 50 & 14100\\
mt3-S11 & 0.51$\pm$0.02 & 12.08$\pm$0.10 & 0.008$\pm$0.001 & 0.2520$\pm$0.0024 & 516 & 26 & 4907\\
mt3-S12 & 0.37$\pm$0.02 & 11.76$\pm$0.08 & 0.019$\pm$0.002 & 0.2479$\pm$0.0031 & 1453 & 54 & 4377\\
mt3-S15 & 0.95$\pm$0.03 & 12.05$\pm$0.11 & 0.012$\pm$0.001 & 0.2524$\pm$0.0022 & 421 & 40 & 5128\\
mt3-S16 & 29.4$\pm$4.6 & 11.84$\pm$0.40 & 0.055$\pm$0.011 & 0.2456$\pm$0.0027 & 33 & 95 & 3514\\
mt4-S1 & 0.45$\pm$0.01 & 11.88$\pm$0.10 & 0.016$\pm$0.001 & 0.2491$\pm$0.0034 & 675 & 30 & 2955\\
mt4-S6 & 1.98$\pm$0.08 & 12.36$\pm$0.24 & 0.018$\pm$0.002 & 0.2508$\pm$0.0035 & 94 & 18 & 1771\\
mt4-S7 & 0.96$\pm$0.07 & 12.06$\pm$0.14 & 0.008$\pm$0.001 & 0.2513$\pm$0.0022 & 302 & 28 & 5285\\
\hline
\multicolumn{8}{c}{Below are data from \cite{MCLIU:Liu2009a} and \cite{MCLIU:Liu2010}}\\
\hline
Mur-S7 & 0.12$\pm$0.01 & 11.96$\pm$0.04 & 0.05$\pm$0.02 & 0.2501$\pm$0.0040 & 6096 & 57 & 1526\\
Mur-S15 & 6.4$\pm$0.6 & 11.37$\pm$0.17 & 0.42$\pm$0.09 & 0.2589$\pm$0.0126 & 60 & 29 & 121\\
mt1-S1 & 0.29$\pm$0.03 & 12.03$\pm$0.08 & 0.14$\pm$0.01 & 0.2554$\pm$0.0092 & 476 & 14 & 150 \\
mt3-S6 & 2.0$\pm$0.2 & 12.12$\pm$0.14 & 0.31$\pm$0.03 & 0.2436$\pm$0.0100 & 113 & 22 & 107 \\
\hline
\end{tabular}}
\label{tab1}
\caption{The Li and B isotopic compositions and concentrations of CM-chondrite spinel-hibonite spherules. [Li], [Be], and [B] are in unit of ppb. Given the counting and systematic errors, the absolute concentrations are probably accurate to 25\%.}
\end{center}
\end{table}

\section{Discussion}
To understand the observed Li and B isotopic compositions in spinel-hibonite spherules, we consider the two irradiation scenarios constructed in \cite{MCLIU:Liu2010}: (1) in-situ irradiation of already-formed hibonite, and (2) irradiation of solar composition gas followed by hibonite formation. In scenario 1, the newly formed hibonite would contain refractory Be, but should be essentially free of the more volatile elements Li and B. Therefore, after irradiation, $^{7}$Li/$^{6}$Li and $^{10}$B/$^{11}$B in hibonite would be expected to be consistent with the spallation production ratios, which are $\sim 1$ and $\sim 0.44$, respectively \cite{MCLIU:Yiou68}. If isotopically normal Li and B were added to the irradiated hibonite at a later stage, the grains would distribute along the mixing trend in a $^{7}$Li/$^{6}$Li-$^{10}$B/$^{11}$B plot, with the two end points being the normal (chondritic) and spallogenic ratios. On the other hand, if the gas of solar composition was irradiated before hibonite formation (scenario 2), the original Li and B isotopic ratios in the gas would have been shifted from chondritic by no more than 0.1\% \cite{MCLIU:Liu2010}. 

\begin{figure}[tbh]
\begin{center}
\includegraphics[width=.65\textwidth]{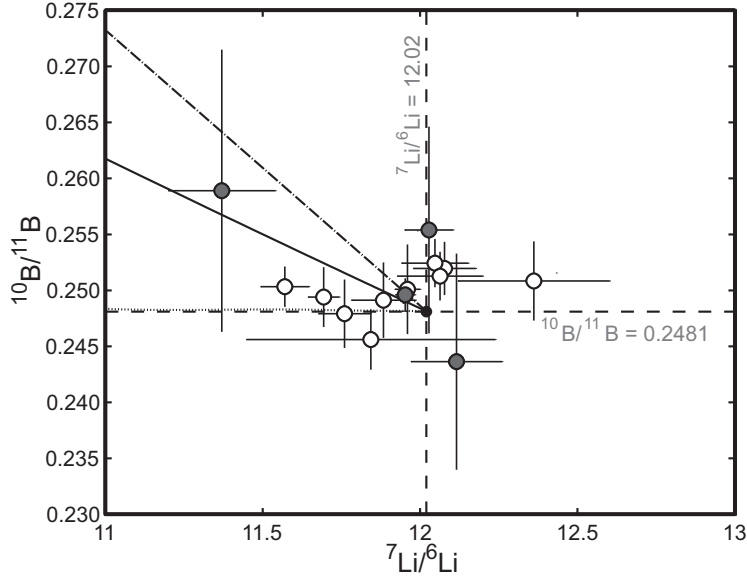}
\caption{The Li and B isotopic compositions of the spinel-hibonites spherules. Open circles represent the new data, whereas the filled ones are literature data from \cite{MCLIU:Liu2009a, MCLIU:Liu2010}. The solid line indicates a hypothetical mixing line between a spallogenic component and an end-member that has chondritic $^{7}$Li/$^{6}$Li (12.02, \cite{MCLIU:Sei2007}) and $^{10}$B/$^{11}$B (0.2481, \cite{MCLIU:Zha96}), and chondritic [B]/[Li] ($\sim$0.49). The gray dash-dotted and dotted lines represent that the contaminating end-member has [B]/[Li] lower (0.2) and higher (106) than chondritic, respectively.}
\label{fig1}
\end{center}
\end{figure}

A $^{7}$Li/$^{6}$Li-$^{10}$B/$^{11}$B diagram for SHIBs measured in this study is shown in Figure~\ref{fig1}. The solid line represents a mixing curve, which is almost straight in the range where the data points distribute, between a ``normal'' reservoir and the spallogenic component in an irradiated hibonite grain, and is calculated as follows:

\begin{displaymath}
\displaystyle\left(\frac{^1X}{^2X}\right)_m = \frac{^1X_s \times f + ^1X_n \times (1-f)}{^2X_s \times f +  ^2X_n \times (1-f)}
\end{displaymath}

\noindent $^1$X and $^2$X represent $^7$Li (or $^{10}$B) and $^6$Li (or $^{11}$B), the subscripts $m$, $s$, and $n$ stand for mixed, spallogenic and normal, respectively, and $f$ is the fraction of spallogenic components in the mixture. The two end-members are assumed to be spallogenic ([Li] = $\sim$2 ppb and [B] = $\sim$6 ppb, when co-produced $^{10}$Be/$^{9}$Be = 1$\times$10$^{-3}$ is achieved in the hibonite target) and chondritic ([Li] = 1.46 ppm and [B] = 0.713 ppm, \cite{MCLIU:Lod2003}). It should be noted that the absolute concentrations of Li and B in the contaminating end-member would not affect the curvature of this mixing curve as long as the abundances of the two elements are kept at their chondritic proportion ([B]/[Li] = 0.49). Four literature data points taken from \cite{MCLIU:Liu2009a, MCLIU:Liu2010} are also included for discussion. Clearly, the data points scatter much more than what would be predicted in the gas-irradiation scenario. Of 14 SHIBs, 7 grains (mt3-S5, mt3-S9, mt3-S10, mt3-S12, mt4-S1, Mur-S7 and Mur-S15) appear to distribute along the solid mixing line between the chondritic and pure spallogenic components, even though supra-chondritic $^{10}$B/$^{11}$B cannot be resolved due to large analytical errors. Isotopically, such a trend is broadly consistent with the prediction of in-situ irradiation of hibonite grains followed by contamination with chondritic Li and B. However, the contaminating end-member must not have exact chondritic [B]/[Li]. Two mixing curves, with the dash-dotted and the dotted lines representing the contaminating end-member's [B]/[Li] = 0.2 and 106, respectively (arbitrarily taken from the lowest and the highest values in the measured SHIBs), are also plotted in Figure~\ref{fig1}. As can be seen in the plot, the mixing curve of [B]/[Li] = 106 passes through essentially all $^{6}$Li-rich grains, indicating that the grains were contaminated with more boron than lithium. This quantitatively explains why the grains that preserve sub-chondritic $^{7}$Li/$^{6}$Li have ``normal'' $^{10}$B/$^{11}$B. Another mechanism that could affect both [B]/[Li] and $^{7}$Li/$^{6}$Li is the diffusion of Li. Because of higher mobility of Li than B and greater diffusivity of $^{6}$Li than $^{7}$Li \cite{MCLIU:Ric2003}, diffusive loss of Li from a grain would definitely increase its [B]/[Li] and $^{7}$Li/$^{6}$Li. One grain in which such a process could have occurred is Mur-S15, because of its low [Li] and high $^{11}$B/$^{6}$Li. From this view, the original (pre-diffusion) $^{7}$Li/$^{6}$Li in Mur-S15 could have been much lower, and it would strongly favor a spallogenic origin. It should be noted that the observed Li and B isotopic compositions could also be explained by isotopic fractionation associated with recondensation of normal gas. However, isotopic or chemical evidence for this process happening to the SHIB grains is still lacking.

\section{Conclusion}
Sub-chondritic $^{7}$Li/$^{6}$Li and supra-chondritic $^{10}$B/$^{11}$B (although less clear) ratios were found in low [Be] spinel-hibonite spherules from the Murchison meteorite. Such isotopic signatures can be best understood in the context of protosolar irradiation of already-formed hibonite solids followed by dilution with isotopically normal Li and B (not necessarily in the relative chondritic abundance). That most $^{6}$Li-rich grains have normal $^{10}$B/$^{11}$B is possibly due to more B contamination than Li. Isotopic fractionation associated with recondensation of normal gas onto the grains could also lead to the observed Li and B isotopic compositions. However, more isotopic and chemical examinations of other elements are needed to prove or disprove this possibility.

\section{Acknowledgments}
The authors thank Andy Davis for providing a Murchison acid residue, Jianhua Wang for keeping the ion probe in top condition, and Marc Chaussidon for inspiring discussion. A constructive review by the referee greatly improved the quality of the paper. This work was supported by NASA Cosmochemistry Program (NNX07AJ71G) and NSC grant (NSC97-2112-M-001-007-MY3).

\end{document}